\documentclass{amsart}
\usepackage{geometry}                		% See geometry.pdf to learn the layout options. There are lots.
\usepackage{graphicx,color}				% Use pdf, png, jpg, or eps§ with pdflatex; use eps in DVI mode
\usepackage{amssymb,amsmath}

\usepackage{hyperref}

\title{To Go Viral}
\author[Ariel Cintr\'{o}n-Arias]{Ariel Cintr\'{o}n-Arias$^{1}$\\ 
$^{1}$Department of Mathematics and Statistics\\
East Tennessee State University\\
Johnson City, TN 37614}

\date{February 13, 2014}							% Activate to display a given date or no date

\begin{document}
\maketitle
%\section{}
%\subsection{}

\begin{abstract}
Mathematical models are validated against empirical data,
while examining potential indicators for an online video that went viral.
We revisit some concepts of infectious disease modeling (e.g. reproductive number) and we
comment on the role of model parameters that interplay in the
spread of innovations.  The dataset employed here provides strong evidence 
that the number of online views is governed by exponential growth patterns, explaining
a common feature of viral videos.  
\end{abstract}

\section{Introduction}

In recent years the phrase {\em ``it went viral"} is commonly
coined to denote popularity and visibility on the internet.
It is almost exclusively reserved for videos, but also applies
to other documents that can be navigated online.

An example of a viral video is Gangnam Style, where Korean singer
Psy leads choreographies with equestrian-inspired moves.  This video 
is enhanced by music that is
remarkably catchy, along with multiple colors in scenarios
induced by an odd sense of humor.  Gangnam Style was uploaded to YouTube\footnote{\url{http://www.youtube.com}}
on July 15, 2012 and since then it has been viewed $1.9\times 10^{9}$ times.
This video is without a doubt the most watched video on YouTube and is also
the first to achieve views on the order of 1 billion \cite{yttrends}.

For mathematicians this type of phenomenon is intrinsically associated with
one of the most eloquent functions in mathematics: the exponential function.  The same
function that connects some of the best-known mathematical constants ($e^{i\pi}=-1$),
and one that is typical in describing phenomena with remarkable growth patterns (such
as bacterial growth).

In this article we discuss how something may {\em go viral}, from
the point of view of mathematical modeling, with particular emphasis on exponential
growth.  We revisit some concepts that are traditionally invoked in models of
infectious diseases, such as reproductive number and final epidemic size, and we comment
on the role played by model parameters that are used to described propagation.

\section{The mathematics of contagion}
Infectious diseases become established, in part, due to 
the mobilization ability of a pathogen that finds its way from host to
host.  One may argue this dynamic process bears resemblance with
way in which people may pass on information, from one person to another.  The most basic mathematical model of
epidemics, developed in 1927 by Kermack and McKendrick \cite{kemc}, 
 was the inspiration, for Daley and Kendall in 1965 \cite{dake65},
 to formulate a simple model
for the spread of unverified information, also known as 
rumors or gossip.  Here we derive an adaption of the latter that applies
to viewership of videos posted online.

Suppose we consider a closed population of constant size, in which individuals
mix homogeneously at random.  In other words, there is an equal chance 
for individuals to encounter one another.  Let us focus on a scenario where 
a new video is posted online and some initial viewers (i.e., early adopters) 
take on the quest of making it available to others.  This hypothetical new video
can be considered a hip innovation such as {\em Gangnam Style}\footnote{Total number of 
views to date 1,891,963,283; \url{http://www.youtube.com/watch?v=9bZkp7q19f0}},
{\em Baby}\footnote{Total number of views to date 976,385,021: \url{http://www.youtube.com/watch?v=kffacxfA7G4}}, 
{\em Call Me Maybe}\footnote{Total number of views to date 527,469,035: \url{http://www.youtube.com/watch?v=fWNaR-rxAic}}, 
and {\em Get Lucky}\footnote{Total number of views to date 47,704,256: \url{http://www.youtube.com/watch?v=h5EofwRzit0}}, 
which so far have cumulative number of views with orders of magnitude between
$10^{7}$ and $10^{9}$.  Under such circumstances the closed
population can be divided into three main groups.  Naive individuals
are those who have never watched the video,
and their size at time $t$ is
denoted by $u(t)$.  Gladwell \cite{gladwell02} 
re-defined mavens as individuals who are ahead of the curve
in identifying fads (likely to trend) and who disseminate them with the aid
of the so-called connectors.  In this model  
mavens and connectors are lumped together in a group of size $v(t)$.
Mavens (also referred to as spreaders)  
are those individuals who have watched the video and are 
actively involved in promoting viewership by means of word of mouth, email, 
social media (e.g. Facebook, Twitter, Google+, Reddit, Tumblr, Pinterest), etc.  The third 
group is denominated stiflers, denoted by $w(t)$, these are 
individuals who stopped being interested in watching the video.

The following nonlinear ordinary differential equations model
the time evolution of these three groups:
		\begin{align}
					\label{naive_eq}
				& \frac{du}{dt} = -b uv\\
					\label{spreader_eq}
				& \frac{dv}{dt} = buv - cv(v+w)\\
					\label{stifler_eq}
				& \frac{dw}{dt} =cv(v+w)\\
					\label{consmass}
				& 1 = u + v + w.
		\end{align}	
The interactions between naive and mavens give
rise to more mavens, as suggested by the term $buv$ in equations (\ref{naive_eq})--(\ref{spreader_eq}),
where the parameter $b>0$ denotes how such interactions take place, and is reminiscent of a rate of transmission.
Moreover, the loss of mavens is assumed to occur proportional to the contacts between individuals
who already know about the video, such that repeated encounters with those who have seen it
may simply discourage spreaders, as they learn it is no longer an innovation.  Spreaders may
feel the video loses appeal once it is perceived as main stream.  The term $cv(v+w)$ denotes
the loss of mavens, where the parameter $c>0$ is called the halting rate.
\begin{figure}[t]
	\begin{center}
		\includegraphics[height=2.5in,width=5in]{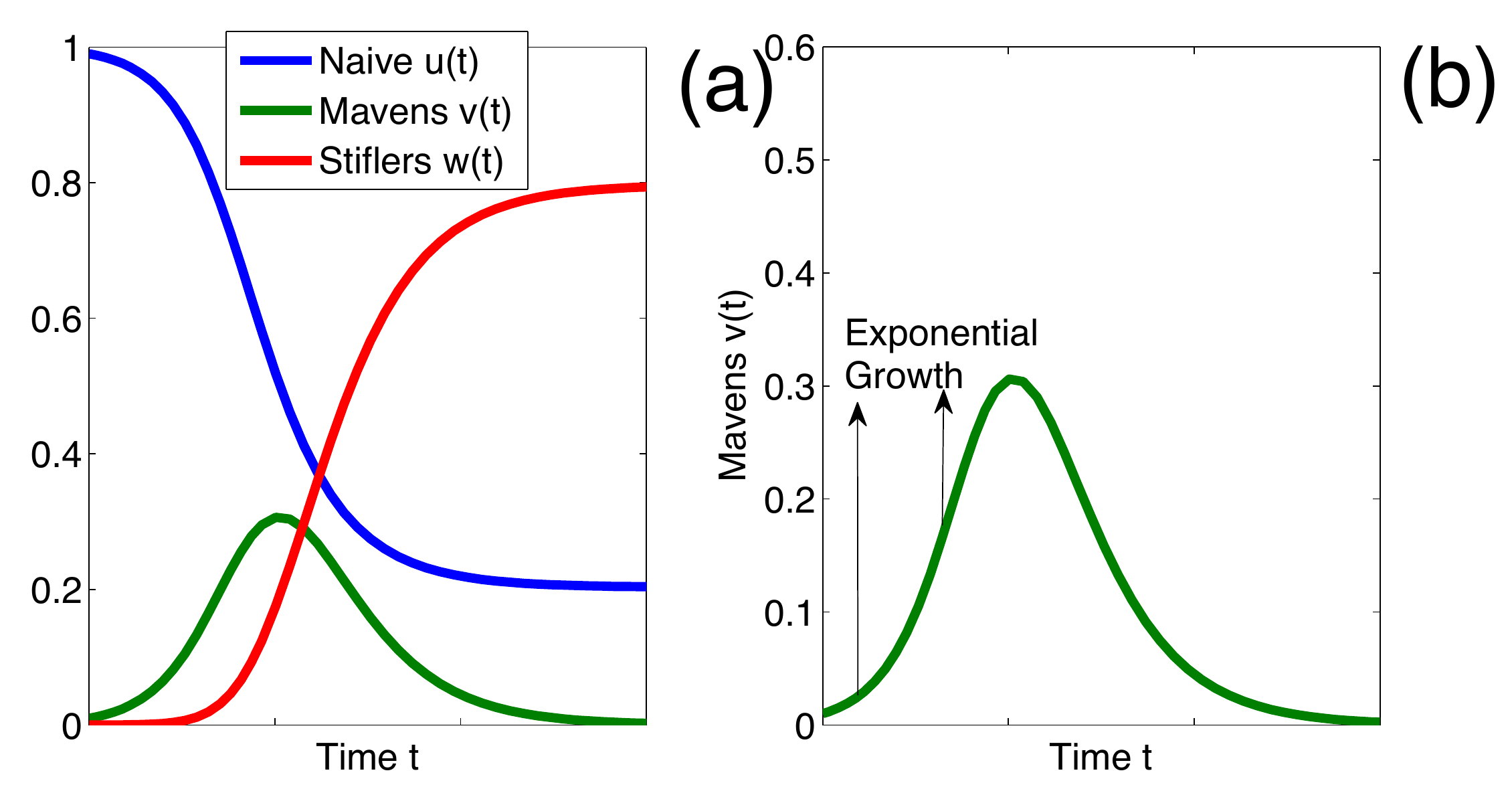}
	\end{center}
	\caption{Panel (a): Numerical solutions of equations (\ref{naive_eq})--(\ref{stifler_eq}) versus time. In Panel (b)
	the curve $v(t)$ is displayed as a function of time $t$, where an exponential
	growth pattern can be observed, 
	for an initial time interval.  Initial conditions and parameter values:  $u(0)=0.99$, $v(0)=0.01$, $w(0)=0$,
	$b=c=0.1$.}
	\label{dk_num}
\end{figure}

Figure \ref{dk_num} portrays numerical solutions of equations (\ref{naive_eq})--(\ref{stifler_eq}).  As 
expected the naive population size is monotonically decreasing,
while the stifler group increases, an implication of negativity and positivity in rates
of change, clearly evidenced in equations (\ref{naive_eq}) and
(\ref{stifler_eq}), respectively.  The mavens population size, $v(t)$, is displayed
in Figure \ref{dk_num}(b) and exhibits the shape of an outbreak curve.

\section{Final size}

The total number of individuals that were at some point mavens
can be quantified by solving a transcendental equation.  First, let us divide
equation (\ref{spreader_eq}) by equation (\ref{naive_eq}) to obtain an expression for
$dv/du$.  Second, using the substitution $v+w = 1-u$ 
we integrate in both sides over the interval $[t_0,t]$, while
assuming $u(t_0)=1$ and $v(t_0)=0$, which reduces to
\begin{equation}\label{dvdu}
	v = (-1)(u-1)\left[1+\frac{c}{b}\right] + \frac{c}{b} \ln u.
\end{equation}

Over the long run the population of mavens vanish while the naive population size approaches a
horizontal asymptote (see Figure \ref{dk_num}(a)).  Thus, let us assume that 
$\lim_{t\rightarrow\infty} v(t) =0$ and $\lim_{t\rightarrow\infty}u(t) = u_{\infty}$.  Taking
the limit as $t\rightarrow\infty$ in both sides of equation (\ref{dvdu}) yields
\begin{equation}\label{fs}
	u_{\infty} = e^{-\mathcal{R}(1-u_{\infty})},
\end{equation}
where $\mathcal{R}=b/c +1$.  Solutions to the transcendental equation (\ref{fs})
in closed form cannot be obtained.  However, numerical approximations
can be computed, provided that a value for the parameter $\mathcal{R}$ is known.
Additionally, it is clear from equation (\ref{fs}) that in the extreme scenario
when $\mathcal{R}$ is sufficiently large, i.e., when $\mathcal{R}\rightarrow\infty$,
then $u_{\infty}\rightarrow0$, implying that the entire naive population eventually
transitions into becoming mavens.  In the literature of infectious 
disease modeling, analogous quantities to $\mathcal{R}$ 
are typically referred to as reproductive numbers  \cite{andmay92,bracas,diek00,het00,satt}.  In classic epidemic 
models with simple dynamics
these reproductive numbers serve as thresholds that separate two main qualitative 
regimes \cite{het00}: an outbreak taking place (whenever $\mathcal{R}>1$)
versus not having enough critical mass to kickoff an epidemic (if $\mathcal{R}<1$).

\begin{figure}[t]
	\begin{center}
		\includegraphics[height=1.8in,width=3in]{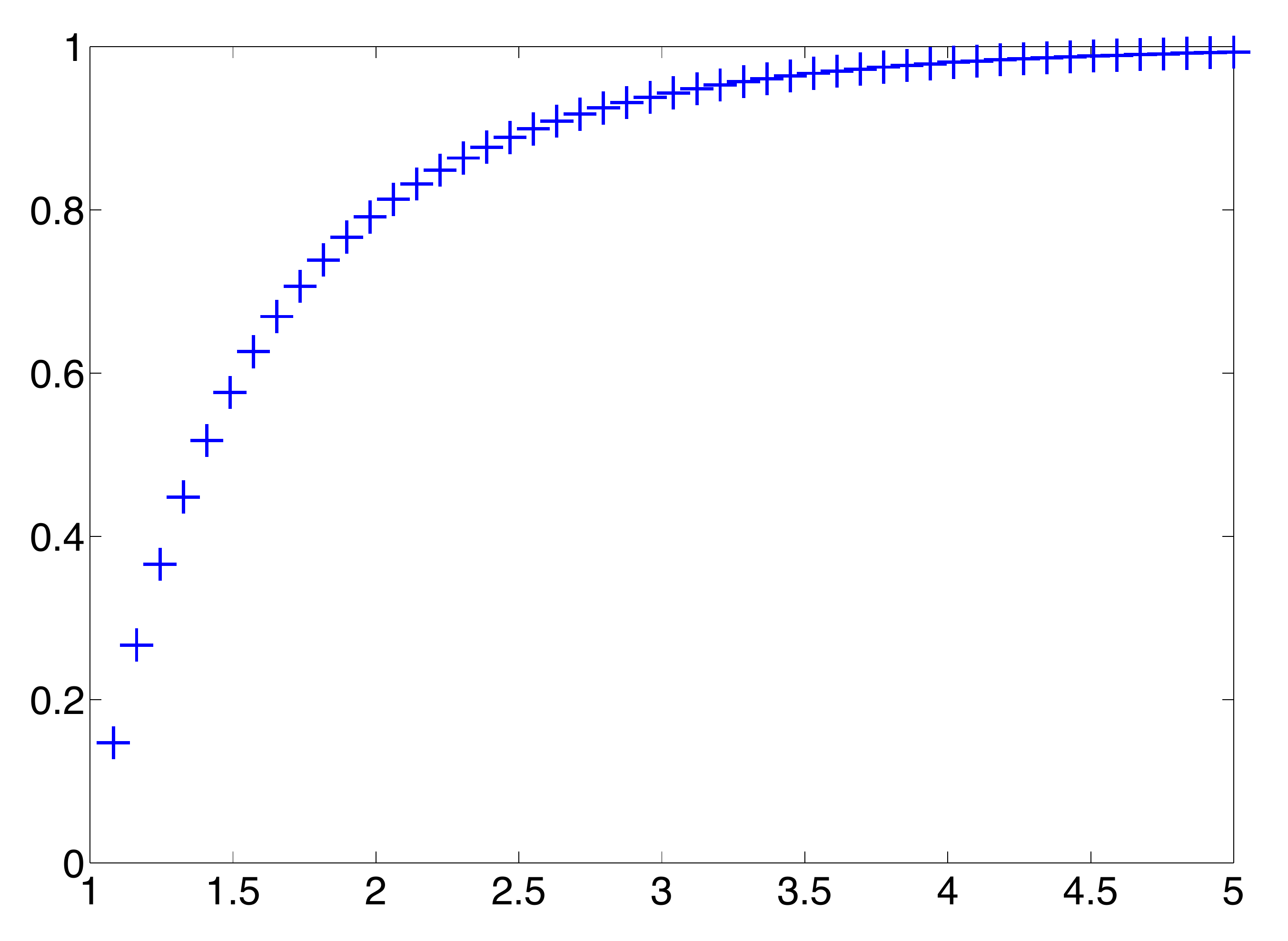}
	\caption{Final size $1-\bar u$ displayed as a function of $\mathcal{R}$, where
	$\bar u$ is a solution to the transcendental equation (\ref{fs}). Fifty equally spaced
	values of $\mathcal{R}$ are used, such that $1\leq \mathcal{R} \leq 5$. Numerical
	solutions were computed by implementing a root-finding algorithm.}
	\label{finalsize}
	\end{center}
\end{figure}

Let $\bar u$ denote a solution to equation (\ref{fs}).  In view of the conservation
of mass assumption given by equation (\ref{consmass}), it then follows that at the end of
an outbreak the quantities $\bar u$ and $1-\bar u$ denote the fractions of those who
never watched the video and those who actively did so, respectively.  It is common
to refer to $1-\bar u$ as the final size of an outbreak. 

Figure \ref{finalsize} portrays numerical solutions of $1-\bar u$ versus $\mathcal{R}$.  For each
value of $\mathcal{R}$ a solution to equation (\ref{fs}) is computed by implementing
an algorithm that finds roots of nonlinear functions\footnote{MATLAB (Mathworks, Inc.) built-in 
function \texttt{fzero} employs a combination of bisection, secant and inverse quadratic interpolation methods.}.  Figure \ref{finalsize} 
illustrates what was previously noted
in relation to how $\bar u$ changes when $\mathcal{R}$ surges.  More specifically, 
as $\mathcal{R}$ increases in value the final size approaches unity, 
i.e., as $\mathcal{R}\rightarrow 5$ then $1-\bar u \rightarrow 1$.

The model defined by equations (\ref{naive_eq})--(\ref{consmass}) differs from its counterpart
epidemic model mainly in the lack of a dynamic threshold.  Because $\mathcal{R} = b/c + 1$ always
exceeds unity an outbreak curve is always guaranteed.  In contrast, for a single-outbreak Susceptible-Infective-Recovered
(SIR) model there is a threshold given by the ratio of transmission to recovery rates, say for example, $\beta/\gamma$, and
if such threshold is below unity then the infective population decays to zero, while when $\beta/\gamma>1$ then an
outbreak curve exists \cite{het00}.  In the next section we carry out a linearization and elaborate more on this feature.

\section{Exponential growth}
\begin{figure}[h]
	\begin{center}
		\includegraphics[height=2.5in,width=5in]{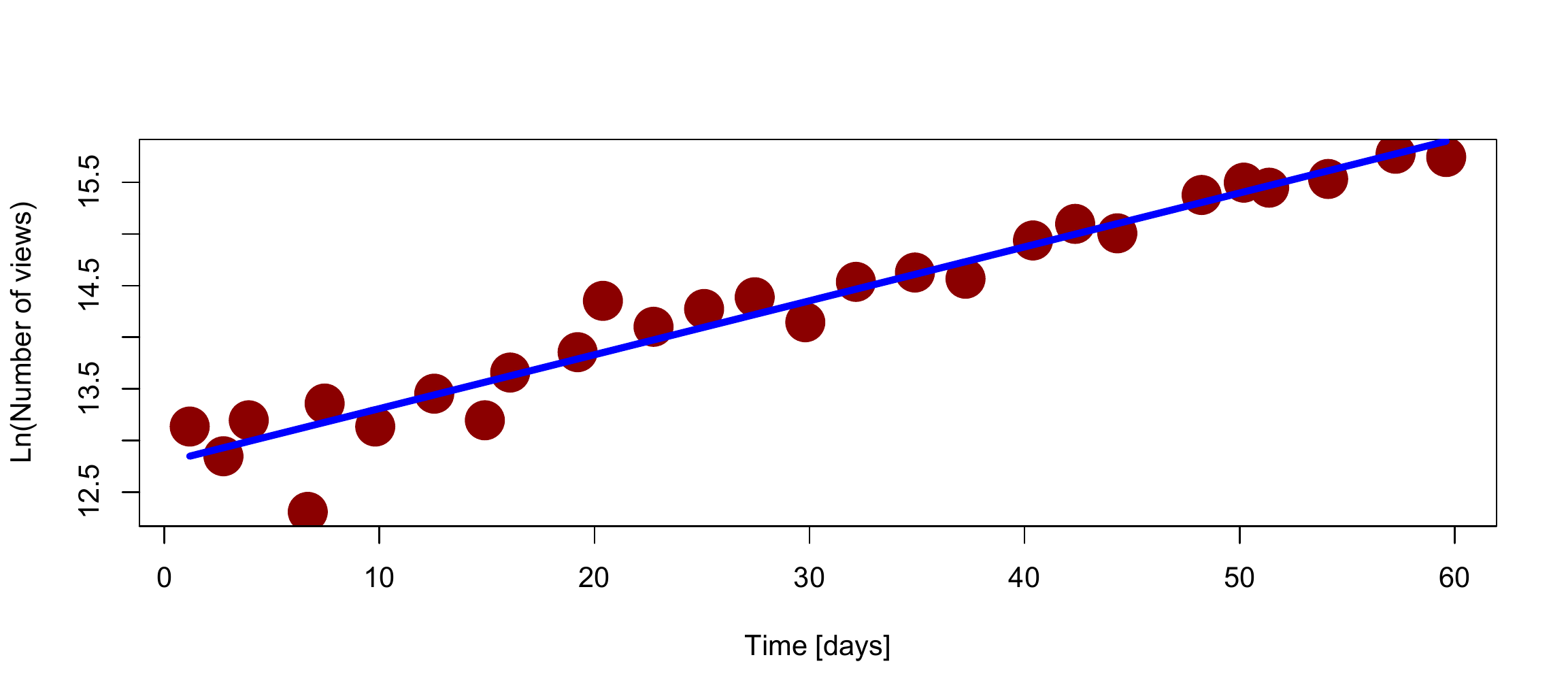}
	\caption{Number of views (circles) versus time and best-fit regression line for
	{\em Gangnam Style} video in YouTube. Data and regression line displayed in logarithmic scale.}
	\label{linreg}
	\end{center}
\end{figure}
				
	The rate of change of the mavens population size can be linearized to
	elucidate understanding on the growth patterns at initial times.  More concretely,
	consider equation (\ref{spreader_eq}) with the substitution $v+ w = 1-u$,
	which is implied by equation (\ref{consmass}):
	\begin{align*}
				& \frac{dv}{dt} = v\left[bu - c(1-u) \right] = cv \left[\left(\frac{b}{c} + 1\right) u -1\right] = f(u,v).
	\end{align*}	
	
	The linearization relies on $\partial f/\partial v$ when $(u,v)\rightarrow(1,0)$ which
	becomes $b$.  Thus, the linearization of $dv/dt$ under the limit $(u,v)\rightarrow(1,0)$
	is given by
	
	\[
		\frac{dv}{dt} \approx bv,
	\]
	which in turn is equivalent to
	\begin{align} \label{expvt}
		&v(t) \approx v(0) e^{bt}.
	\end{align}
	
	Equation (\ref{expvt}) is an approximation in the so-called {\em invasion limit}, i.e., $(u,v)\rightarrow(1,0)$,
	where the word invasion finds its roots in the theoretical ecology literature with the context 
	of having a species invade another species \cite{cas01,edel88,kot01,murr,otto11,thieme03}.  Here invasion 
	refers to a population that is initially
	made of all naive individuals having only an infinitesimal presence of mavens.

	The equivalent calculation leading to equation (\ref{expvt}) for an SIR model (see Appendix)
	yields $\beta-\gamma$, instead of $b$.  Thus, the faith of the infective population is determined 
	by the positivity or negativity of $\beta-\gamma$.  More specifically, when $\beta/\gamma > 1$
	the infective population initially grows exponentially, otherwise it is destined to
	exponential decay.  The fact that $b>0$ implies a guaranteed exponential take off is
	one the main distinctions between a single-outbreak SIR model and the model defined by equations
	(\ref{naive_eq})--(\ref{consmass}).

	Let us suppose the number of mavens at time $t$ is given by $v(t)\approx x(t) = A e^{bt}$, where the parameter $A$
	denotes the initial condition, i.e., $A=v(0)$.  Whenever a temporal dataset is available then it is straightforward
	to estimate the parameters $A$ and $b$.  It is common to transform the data and model using a natural
	logarithm transformation. The model becomes
	\[	
		y(t) = \ln x(t) = \alpha + bt,
	\]  
	where $\alpha = \ln A$.  A log-transformed dataset $\{ (t_1,\ln x_1),(t_2,\ln x_2),\dots,(t_n,\ln x_n) \}$ can used
	to compute point-estimates, denoted as $\hat \alpha$ and $\hat b$,  
	of the parameters $\alpha$ and $b$ while applying
	linear regression formulas (see \cite{seier11} and Appendix for details).

	Viewership data for the music video {\em Gangnam Style}, that is, 
	data on the number of views versus time, were available in the blog known as YouTube 
	Trends \cite{yttrends}.  These temporal observations are displayed in Figure \ref{linreg} (circles) in logarithmic
	scale.  Moreover, the best-fit regression line $y(t) = \hat \alpha + \hat b t$ also appears in Figure \ref{linreg}.  Parameter estimates
	within one standard error\footnote{The bulit-in function \texttt{nls} of the open-source statistical package R
	was implemented to compute standard errors.} are:
	\begin{align}
		& \hat b = 0.0522  \pm 0.0026 \\
		& \hat \alpha = 12.7856  \pm 0.0883 
	\end{align}	
	The Pearson correlation coefficient (see \cite{seier11} and Appendix)
	for the data displayed in Figure \ref{linreg} equals $0.9700$, implying a strong correlation
	between the $\ln x_i$'s and $t_i$'s.  Moreover, the coefficient of determination
	is then $0.9700^2 = 0.9409$, meaning that 94\% of the variability in the data can
	be explained by a linear model \cite{seier11}, in other words, a linear model is justifiably well 
	suited.  The strong correlation in the logarithmic scale translates into solid evidence
	of exponential growth patterns for the viewership data, in the particular case considered here.  In other words, 
	there is strong evidence that the number of views for {\em Gangnam Style} obeys exponential growth.  It is precisely this 
	feature what is commonly associated with the phrase {\em ``it went viral"}.

\section{The most influential parameter}

\begin{figure}[t]
	\begin{center}
		\includegraphics[height=3.5in,width=5.5in]{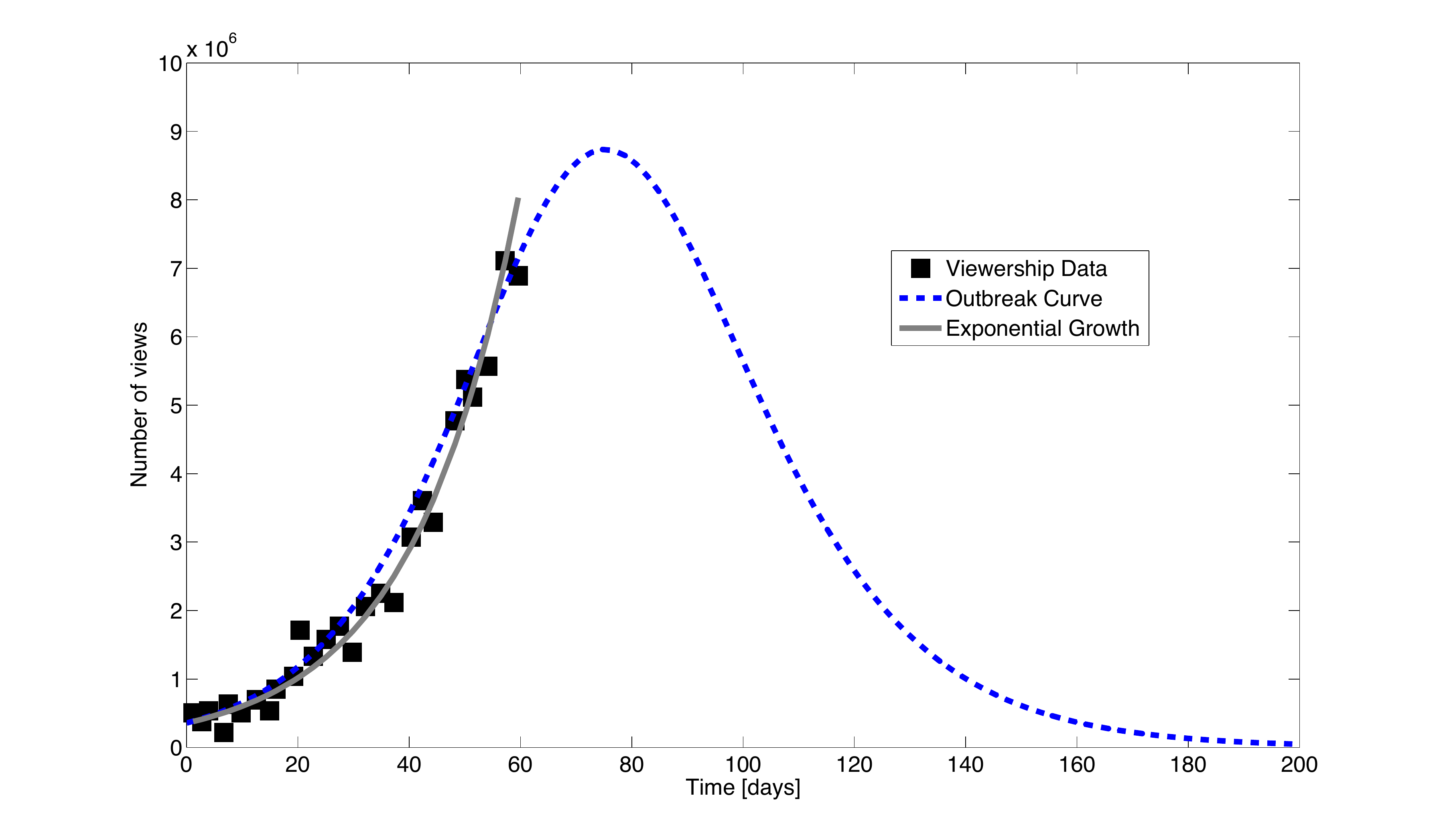}
	\end{center}
	\caption{Longitudinal observations on the number of views of {\em Gangnam Style} are depicted with 
	squares.  Best-fit exponential model, defined in equation (\ref{expvt}),
	is displayed as a gray solid curve.  Best-fit outbreak curve, $v(t)$ in equations (\ref{naive_eq})--(\ref{stifler_eq})
	scaled by a total population size $N= 7.0\times10^{7}$, appears as a blue dashed curve.	
	Initial conditions: $v(0) = 3.6\times10^{5}/N$, $u(0)=1-v(0)$, $w(0)=0$.  
	Model parameter estimates and standard errors are given in the text.} 
	\label{olsestimates}
\end{figure}

The same dataset that was used in the previous section, for a linear regression analysis, is now employed
to validate the mathematical model defined by equations (\ref{naive_eq})--(\ref{stifler_eq}).  This model is validated against
empirical data by applying ordinary least squares (OLS) methods for inverse problems
(see \cite{bankstran,ca09} and references therein for additional details).

The model parameters are the transmission rate $b$ and the halting rate $c$, and 
their OLS estimates are
denoted by $\hat b$ and $\hat c$.  The numerical solution to equations (\ref{naive_eq})--(\ref{stifler_eq})
obtained with the parameter estimates is usually denominated the best-fit solution.  Figure \ref{olsestimates} displays
the temporal data (squares) and the best-fit solution (dashed curve).  The OLS parameter estimates within one 
standard error are\footnote{Numerical solutions of OLS inverse problems can obtained with implementations in MATLAB
that include the Global Optimization toolbox.  Built-in functions \texttt{ga} and \texttt{patternsearch} were employed here.}
\begin{align}
	& \hat b = 0.0623 \pm 0.0020 \\
	& \hat c =  0.2102 \pm 0.0413.
\end{align}

Because data is given in raw quantities and not in percentages or fractions, as the mathematical model was
formulated (see equation (\ref{consmass})), it was necessary to also estimate a total population size, which was used as a scaling factor.  For
this quantity, only a point estimate (without uncertainty bounds) was computed, namely, $N =  7.0359 \times10^{7}$.

For the sake of comparison, the exponential model defined in equation (\ref{expvt}) was also fitted to the longitudinal dataset.
The best-fit exponential solution to equation (\ref{expvt}) appears displayed in Figure \ref{olsestimates} as a solid 
curve.  The exponential model has two parameters: $A=v(0)$ and $b$.  Estimates plus minus one standard error for these parameters
are given by
\begin{align}
	& \hat A =  4.3590 \times 10^{5}   \pm 4.1370 \times 10^{4} \\
	& \hat b= 4.7750 \times 10^{-2}  \pm 1.8510 \times 10^{-3}. 
\end{align}

Whenever mathematical models are validated against empirical data, it is customary to address
the role that parameters play in state variables.  In other words, how variability in model
parameters manifest in the output of the model.  One can rule dependence on model parameters by computing
sensitivity functions.  For example, let us consider equation (\ref{expvt}), where we write $x(t) = A e^{bt}$.  The partial
derivatives 
$\partial x/\partial A$ and $\partial x/ \partial b$ are denominated traditional sensitivity functions \cite{banks07}.  They
are time functions addressing the rate of change with respect to parameter variability.  A re-scaled
version of these functions is useful in circumventing issues of magnitude and scale.  Define
\[	
	\nu_A(t) = \frac{A}{x(t)} \frac{\partial x}{\partial A}(t) \equiv 1
\]
and
\[
	\nu_b(t)= \frac{b}{x(t)} \frac{\partial x}{\partial b}(t) =bt.
\]

\begin{figure}[t]
	\begin{center}
		\includegraphics[height=3.5in,width=5.5in]{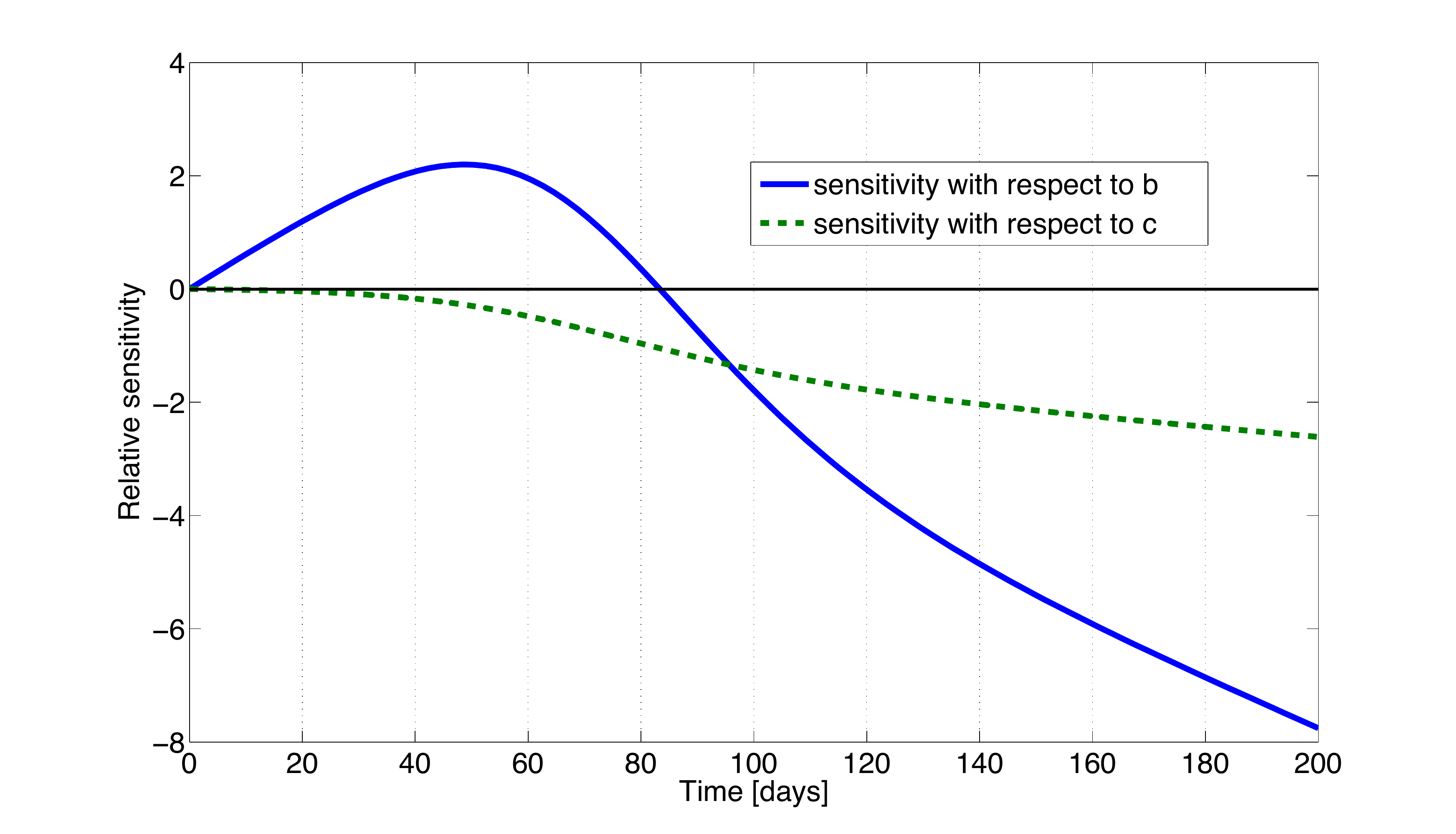}
	\end{center}
	\caption{Relative sensitivity functions with respect to model parameters $b$ and $c$, 
	versus time.} 
	\label{relsen}
\end{figure}
The functions $\nu_A(t)$ and $\nu_b(t)$ are called relative sensitivity functions.  For the exponential
growth model, equation (\ref{expvt}),
both of these functions are linear, one of them with zero slope while the other one has slope $b>0$.
On one hand, when the relative sensitivity functions approach zero one concludes the rate of change with respect
to that parameter vanishes, meaning that no dependence on that parameter exists.  On the other hand,
whenever the relative sensitivity functions are away from zero, then monotonic dependence on the parameters
can be ruled in consistency with the sign: negativity would imply decreasing dependence, while
positivity would suggest increasing dependence.  More specifically, the exponential growth model
has $\nu_A(t)>0$ and $\nu_b(t) >0$ implying that $x(t)$ increases when $A$ and $b$ increase.  Furthermore,
if $t>1/b$ then $\nu_b(t) > \nu_A(t)$, meaning that $b$ is the most influential parameter in $x(t)$.

Relative sensitivity functions cannot be computed in analytic form for the state variables of
equations (\ref{naive_eq})--(\ref{stifler_eq}).  Because this is a nonlinear system for which
closed form solutions are not available.  Instead numerical solutions can be computed
by using the parameter estimates to solve an extended system that includes auxiliary equations.  
The latter equations are known as the {\em forward sensitivity equations}, where the 
state variables of the original system act as time-dependent coefficients for a linear (auxiliary) system, 
with unknowns equal to the partial derivatives of the state variables with respect to model parameters.
For additional details in numerical solutions of forward sensitivity equations, the reader is referred to 
\cite{banks07,bankstran,ca09} and references therein.

Let us define the relative sensitivity functions for the mavens population size $v(t)$, 
from equations (\ref{naive_eq})--(\ref{stifler_eq}), as follows:

\[
	\phi_{b}(t) = \frac{b}{v(t)} \frac{\partial v}{\partial b}(t)
\]
and
\[
	\phi_{c}(t) = \frac{c}{v(t)} \frac{\partial v}{\partial c},
\]
where $\partial v/\partial b$ and $\partial v/\partial c$ are computed numerically.  Figure \ref{relsen}
displays $\phi_b(t)$ (solid curve) and $\phi_c(t)$ (dashed curve) versus $t$.  It is clear from 
Figure \ref{relsen} that for $t\in[0,30]$ we have that $\phi_c(t)\approx 0$, implying the 
halting rate $c$ does not influence $v(t)$ over this initial time interval.  The model parameters
play an expected role for $30\leq t \leq 83$
where $v(t)$ increases with the transmission rate 
$b$ (because $\phi_b(t)>0$) and $v(t)$ decreases with the halting rate $c$ (due to $\phi_c(t)<0$).  
Surprisingly, the role of parameter $b$ is reversed when $t>83$, in the sense that $\phi_b(t)<0$.  Furthermore,
towards the end of the outbreak, as $t\rightarrow200$, we have that $\phi_b(t)<\phi_c(t)<0$,
suggesting that $b$ is more influential than $c$ in the ending phase, because $\phi_b(t)$
is farthest away from zero.  In summary, the transmission rate $b$ is the most influential parameter.

We also note that having $\phi_c(t)\approx 0$ and $\phi_b(t)>0$ for $0\leq t \leq 30$, also
confirms that during this time interval the curve $v(t)$ is only influenced
by the transmission parameter $b$, but it is also heavily dominated by exponential growth.
The latter is confirmed in Figure \ref{olsestimates} by comparing the best-fit curves (solid versus dashed),
where reasonable agreement with the pattern displayed by the longitudinal data is 
observed for $t\in[0,30]$.  Temporal patterns of exponential growth are, to the 
best of our knowledge, footprints of going viral.

\section{Final remarks}

The website known as YouTube was launched in early 2005 with the innovative goal of providing a user-friendly
means to share videos.  Over nearly a decade, it is clear this website is revolutionizing online video streaming.  A 
technology that is replacing video stores by online rentals (e.g. Amazon Instant Video, Netflix).

In the context of our contemporary era of social networking sites, powered by 
digital media, traditional terms in epidemiology of communicable diseases
(such as transmissibility, reproductive number, final size, etc)
find new meanings, when viewed through the lens of marketing and digital strategy.  In reverse 
sense of what we would aim with infectious diseases (most efforts target eradication),
digital media would intend to establish trends.  More importantly, it is of great interest
to determine what factors (e.g. age group, demographics, education level, gender, etc) 
would favor the likelihood of establishment.  Mathematical, statistical and computational methods
(e.g. data science and predictive modeling) are merging together to 
address questions in the latter context, however the challenge remains paramount.    

In this paper, we have revisited a mathematical model
proposed by Daley and Kendall \cite{dake65}, in the context
of rumor propagation by word of mouth.  Most of the active 
development of theory centered around this model relates to
stochastic modeling \cite{isham10,lefpic,matho,mor2,pea,sud1,wat,zan1}, including
explorations of social landscapes equivalent
to complex heterogeneous networks.  However, to the best of
our knowledge, most of this theoretical work remains yet to be validated
against empirical data.   

In this study we were successful in validating Daley-Kendall's model
against longitudinal observations.  Nevertheless, the parameter estimates must
be considered with a degree a caution, because they only constitute
first order approximations conditional on the dataset.  The longitudinal counts
on the number of views of Gangnam Style are aggregate, as they inevitably
include repeated counts from individuals who watched the video more than once.
The model parameter estimates and their uncertainty bounds, can only
be accepted by making the simplifying assumption that the viewership data
approximate the actual number of mavens at each time point. 

More important than the estimated parameter values is how the process
of dissemination is readily verified with simple mathematical models.  A contagion model,
equations (\ref{naive_eq})--(\ref{consmass}), can successfully describe the surge and fall 
of mavens, as it is clearly seen in Figure \ref{olsestimates}.  A better dataset to validate 
a contagion model would include observations at the peak and in the downfall phase
of the outbreak curve, not just at the beginning.   

A central point for our analysis here is to articulate in mathematical terms
what it means when a video {\em ``went viral"}.  Figures \ref{linreg} \& \ref{olsestimates}
confirm that to go viral is the equivalent of exponential growth.  More specifically, 
 we find that strong positive correlation (with a Pearson coefficient of 0.97) 
 in logarithmic scale is an indicator of exponentially growing patterns.  The transmission rate
 $b$ denotes the ability to promote viewership,     
and it is the most influential parameter for both the contagion and the exponential models.  
One of the features of exponentially growing processes is that they can easily
climb up orders of magnitude within a short time window, a feature that is well documented
for viral videos.  For example, Gangnam Style has views jumping from order $10^{5}$
(on day 0) to $10^{6}$ (on day 20).

On a closing note, mathematicians can only hope that some their publications
in peer-reviewed journals {\em ``went viral"}.  Perhaps the closest metric of such degree
of visibility is the number of citations.  This author would be thrilled if any of his 
publications are ever cited with order of magnitude $10^2$.  But yet it is still a long
way to go, because his most cited paper has to date 
71 citations (according to Web of Knowledge\footnote{\url{http://wokinfo.com}}), and when 
considered over a period of 8 years, it leaves much to be desired.

\section*{Acknowledgments}
The author thanks Tara Gancos Crawford, Rachel Fovargue, Amanda Traud, Jian Zhang,
Virginia Ingram, 
and the National Institute for Mathematical and Biological Synthesis, for helpful discussions
on this topic.

\section{Appendix}
	{\bf SIR Model.}  A single outbreak Susceptible-Infective-Recovered model is defined by the nonlinear
	system of equations (see \cite{het00} and references therein for additional details):
	\begin{align*}
		& \frac{ds}{dt} = -\beta si\\
		& \frac{di}{dt} = \beta si -  \gamma i \\
		& \frac{dr}{dt} =	\gamma i\\
		& 1 = s+i+r,
	\end{align*}
	where the epidemic parameters are $\beta$ and $\gamma$.  A linearization of the infectives equation, around
	$s=1$ and $i=0$, yields
	\[
		\frac{di}{dt} \approx (\beta-\gamma)i,
	\]
	which implies, 
	\[i(t)\approx i(0)e^{(\beta-\gamma)t}.\]

	{\bf Linear Regression.}  Given the observations $(t_1,\ln x_1),(t_2,\ln x_2),\dots,(t_n,\ln x_n)$,
	then point-estimates for the parameters of the linear model
		\[
			y(t) = \ln x(t) = \alpha + b t,
		\]
	are computed as follows \cite{seier11}:	
	\begin{align*}
		& \hat b = \frac{\sum_{i=1}^{n}\left(t_i-\bar t \ \right)\left(\ln x_i -\bar y \right)}{\sum_{i=1}^{n}\left(t_i-\bar t \ \right)^2}\\
		&  \hat \alpha = \bar y - \hat b \bar t 
	\end{align*}
	where
	\[
		\bar t = \frac{1}{n}\sum_{i=1}^{n}t_i
	\]
	and
	\[
		\bar y = \frac{1}{n}\sum_{i=1}^{n} \ln x_i. 
	\]		

	The sample standard deviations for the $t_i$'s and $\ln x_i$'s are denoted by $s_t$ and $s_y$, respectively, 
	and are computed using the sample averages $\bar t$ and $\bar y$, namely \cite{seier11}:
	\begin{align*}
		&s_t = \sqrt{\frac{\sum_{i=1}^{n}(t_i-\bar t)^2}{n-1}}
		\hspace{.5in}
		 \mbox{and} 
		&s_y=\sqrt{\frac{\sum_{i=1}^{n}(\ln x_i-\bar y)^2}{n-1}}.
	\end{align*}	 
	
	The Pearson correlation coefficient is defined in terms of the sample average and sample standard deviation \cite{seier11}:
	\[
		\frac{1}{n-1}\sum_{i=1}^{n} \left(\frac{t_i-\bar t}{s_{t}}\right)  \left(\frac{\ln x_i-\bar y}{s_{y}}\right).
	\]

\end{document}